\documentclass{ifacconf}[a4paper]
\usepackage{graphicx}      % include this line if your document contains figures
\usepackage{natbib}        % required for bibliography

\usepackage{xurl}                  %ADDED
\usepackage{amsmath,amssymb}      %ADDED
\usepackage{tabularx}             %ADDED
\usepackage{makecell}             %ADDED 
\usepackage{caption}   
\usepackage{microtype}

\begin{document}
\begin{frontmatter}

\title{Hierarchical Control via MPC-RL for  Multi-Timescale Battery Systems}%\thanksref{footnoteinfo}} 

%\thanks[footnoteinfo]{Sponsor and financial support acknowledgment
%goes here. Paper titles should be written in uppercase and lowercase
%letters, not all uppercase.}

\author[1]{Rasa Pourjam}
\author[1]{Ehecatl Antonio del R\'io Chanona}
\author[1,2]{Paulina Quintanilla}

\address[1]{Sargent Centre for Process Systems Engineering, Department of Chemical Engineering, Imperial College London, London SW7 2AZ, United Kingdom
(e-mail: a.del-rio-chanona@imperial.ac.uk)}

\address[2]{Department of Chemical Engineering, University College London, London WC1E 6BT, United Kingdom
(e-mail: p.quintanilla@ucl.ac.uk)}

\begin{abstract}               
% Abstract of 50--100 words
Multi-timescale systems present a fundamental challenge, where fast operational decisions must coexist with long-horizon sustainability targets. In this work, we proposed a new hierarchical control framework via Model Predictive Control (MPC) and Reinforcement Learning (RL) to separate decision-making on two distinct timescales. The high-level MPC optimizes long-horizon setpoints at the slow dynamic and on a fast timescale, a low-level pretrained RL agent tracks these setpoints in real time to maximize short-term objectives. RL is introduced to learn nonlinear control policies, without relying on model linearizations or requiring the heavy online computation from solving repeated optimal control problems. The framework is applied to a Battery Energy Storage System (BESS) operating in frequency regulation markets to balance fast profit opportunities (seconds) and slow battery degradation (weeks to months). %The control design consists of: (1) MPC data generation, (2) Behavior Cloning, (3) degradation-aware RL, and (4) degradation-unaware RL for setpoint tracking, before integration with a high-level MPC in a two-layer hierarchy. 
%The control design consists of: (1) MPC data generation, (2) Behavior Cloning, (3) degradation-aware RL to learn long-horizon, battery-health-conscious behavior and generate safe reference setpoints, and (4) degradation-unaware RL fine-tuned from the previous stage to perform fast setpoint tracking at runtime, before integration with a high-level MPC in a two-layer hierarchy. 
The design employs a degradation-aware RL agent trained offline to generate safe long-horizon setpoints, and a degradation-unaware agent fine-tuned from it for fast runtime setpoint tracking. %The resulting controller was evaluated on three BESS chemistries to assess its generalizability. 
Compared to MPC baselines, the proposed approach successfully extends battery lifetime by 84\% and increases operational profit by 34\%.
\end{abstract}


\end{frontmatter}
%===============================================================================

\section{Introduction}

BESS are essential for the integration of renewable energy to ensure grid stability through the frequency regulation market (FR) \citep{hambissa2024analysis}. These systems store surplus renewable generation and release it later when demand exceeds supply. By rapidly charging and discharging, BESS smooths the variability of renewable energy output and helps maintain the generation–load balance enough to satisfy the frequency limits of the grid\citep{ma2023life}. These fast responses generate revenue opportunities but accelerate capacity degradation, reducing battery lifespan and reducing the return on initial investment \citep{vykhodtsev2022review}.This problem creates a trade-off between short-term revenue and long-term battery health, making it a multi-time-scale system. Their operation is crucial for maintaining long-term economic viability \citep{reniers2021unlocking}.

Recent studies show that including battery degradation in the control strategy can increase expected profits by 12\% to 46\%, depending on the control strategy and the chosen end-of-life limit (EOL) \citep{wankmuller2017impact}. To put this into practice, the control framework needs a degradation model that captures how aging limits capacity and how operating choices influence the rate of aging \citep{zagorowska2020survey}. Effective control must balance short-term revenue with long-term degradation \citep{cao2020multiscale}.

Conventional rule-based or single-layer MPC schemes optimize short-term profit but overlook degradation effects, and requires solving thousands of large-scale nonlinear programs at a fast timescale. Recent studies have introduced degradation-aware MPC and RL frameworks to improve BESS operation under uncertainty \citep{li2022deep,cao2020multiscale}. RL can also handle nonlinear process dynamics without relying on model linearizations and avoids repeated large optimal-control solves \citep{sachio2022integrating}. However, purely RL-based methods have been tested only on a single type of BESS; robustness to measurement noise and generalization across chemistries remain unaddressed. %However, purely RL-based methods have been tested only on a single type of BESS with ideal state measurements. Stability concerns also matter in BESS, because safety limits and measurement noise can cause inconsistent control actions.

This work introduces a new hierarchical MPC–RL control framework for a multi-time-scale system. The proposed architecture separates decision-making on two distinct timescales. The high-level MPC optimizes long-horizon, degradation-aware setpoints at the slow time-scale. On a fast timescale, a low-level pretrained RL agent tracks these setpoints in real time, learning to exploit short-term fluctuations in FR signals to maximize revenue while respecting the MPC-generated safety envelope. We evaluated three different battery systems to examine the ability of framework to generalize between different characteristics of the system.

\section{Methodology}

\begin{figure}[h!]
    \centering
%    \vspace{-10pt}
    \includegraphics[width=0.9\linewidth]{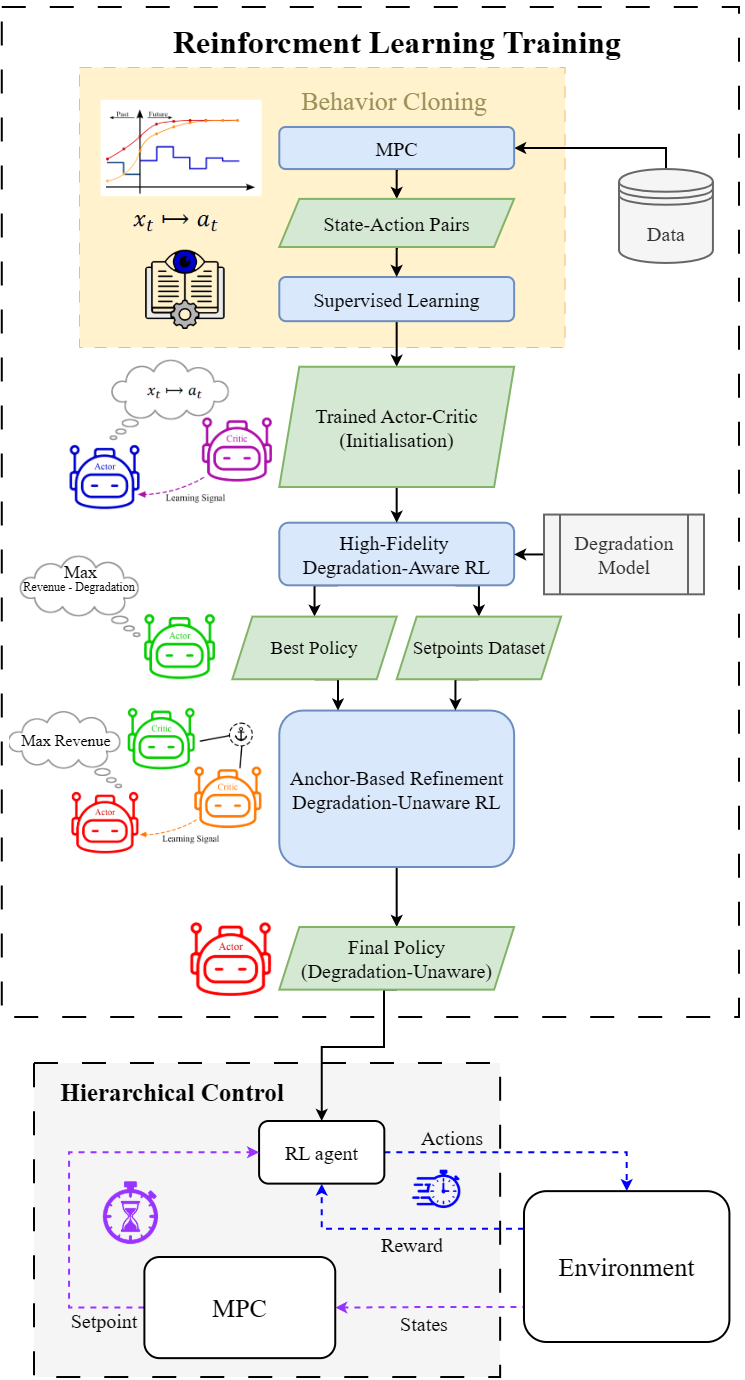}
    \caption{Framework of the proposed methodology.}% RL is pretrained under four stages before integration into the hierarchical control.}
 %   \vspace{-5pt}
    \label{Fig. 1}

\end{figure}

The proposed framework employs a multi-stage training pipeline to robustly initialize the RL agent before hierarchical MPC integration. A behavior-cloning dataset from MPC rollouts first initializes the actor-critic via supervised learning. This policy is refined in a high-fidelity environment (differential and algebraic equation (DAE) system) by coupling a single-particle model (SPM) \citep{santhanagopalan2006review} with solid-electrolyte interphase (SEI) degradation kinetics to learn a degradation-aware policy that explicitly accounts for capacity fade. To enable real-time 2-second execution, the degradation-aware policy is distilled via anchor-based refinement into a degradation-unaware agent that tracks MPC setpoints without requiring online degradation states. Computationally, the low-fidelity linear energy balance model is reserved for fast MPC scheduling, while the DAE-based environment is used exclusively for offline RL training and evaluation. The framework is parameterized for multiple chemistries, including LiFePO$_4$ (LFP), nickel–manganese–cobalt (NMC) and nickel-cobalt-aluminum (NCA) \citep{forman2012genetic, safari2011modeling}. These chemistries differ in electrochemical, transport, and SEI-degradation parameters. We test controller generalizability by applying the same MPC-RL workflow across systems with different dynamic and aging characteristics. The complete workflow is illustrated in Fig.~\ref{Fig. 1}.

\subsection{Data Structure}
%\vspace{-10pt}
We use the Pennsylvania-New Jersey-Maryland (PJM) Interconnection dataset provided in \citep{cao2020multiscale}. The FR signal is normalized, with a two-second resolution and a range of $[-1,+1]$.  When the signal is positive, the battery is asked to send power to the grid and discharge; when it is negative, the battery should charge from the grid. The dataset also contains hourly day-ahead electricity prices and incentive payments associated with providing FR capacity.

\subsection{Decision Variables}
%\vspace{-10pt}
In the FR market participation problem, the control variables are defined on an hourly schedule. The battery commits a FR capacity for hour $t$. During that hour, the independent system operator (ISO) may call any proportion of this band every two seconds through the FR signal, and the asset is paid the hourly FR incentive price. We include an adjustable on-site load, served from the battery to help manage state of charge (SoC), which is the fraction of usable energy currently stored in the battery. The load is assumed to remain constant on an hourly basis and does not earn direct revenue. 
%\vspace{-10pt}
        
\subsection{MPC}
\label{sec:mpc}

%\vspace{-10pt}
MPC is applied in a low-fidelity environment to schedule battery participation in the FR market. At each decision point  $t$, the MPC solves an optimization over a prediction horizon of $N=4$ hours. Each hour \( k \in \{t, \ldots, t+N-1\} \) is discretized into $\mathcal{S}=1800$ steps of 2 seconds, indexed by $s \in \{1,\ldots,\mathcal{S}\}$. The MPC optimization problem at time $t$ is as follows:
\vspace{-2pt}
\begin{equation}
\max_{\{F_k, O_k, L_k, P_{k,s}, E_{k,s}\}} \sum_{k=t}^{t+N-1} \left( \pi_k^f F_k - \pi_k^e O_k \right)
\end{equation}

\begin{align}
\text{s.t. } & E_{k,s+1} = E_{k,s} + \Delta t P_{k,s}, \quad \forall k, \, s = 1,\ldots,\mathcal{S}, && \tag{1.1} \\[6pt]
& \underline{P} \leq P_{k,S} \leq \overline{P}, && \tag{1.2} \\[6pt]
& (1-c_f)\lambda_l E_{max} \leq E_{k,s} \leq (1-c_f)\lambda_u E_{max}, && \tag{1.3} \\[6pt]
& P_{k,s} = \alpha_{k,s}F_k + O_k - L_k, && \tag{1.4} \\[6pt]
& F_k \geq 0, \quad O_k \geq 0, \quad L_k \geq 0, && \tag{1.5} \\[6pt]
& (1-c_f) \epsilon_l E_{max} \leq E_{t+N,1} \leq (1-c_f) \epsilon_u E_{max} && \tag{1.6}
\end{align}

\noindent{where $\pi_k^f, \pi_k^e$ are the FR incentive price and electricity price at hour $k$ (hourly basis), $F_k$ is the FR capacity committed to the grid at hour $k$, $O_k$ is the power purchased from the grid at hour $k$, $L_k$ is the adjustable load at hour $k$, $P_{k,s}$ is the net charging/discharging power within each hour, $E_{k,s}$ is the SoC at step $s$, $\alpha_{k,s}$ is the normalized FR signal $\in [-1,1]$, $E_{max}$ is the nominal battery capacity, reduced by $(1 - c_f)$ to reflect the remaining usable energy, $\lambda_l, \lambda_u, \epsilon_l, \epsilon_u \in (0,1)$ are the margins to ensure energy reserves, and $ c_f$ is the cumulative capacity fade.
% \begin{itemize}
%     \item[] $-\ \pi_k^f, \pi_k^e$: FR incentive price and electricity price at hour $k$ (hourly basis).
%     \item[] $-\ F_k$: FR capacity committed to the grid at hour $k$.
%     \item[] $-\ O_k$: power purchased from the grid at hour $k$.
%     \item[] $-\ L_k$: adjustable load at hour $k$.
%     \item[] $-\ P_{k,s}$: net charging/discharging power within each hour.
%     \item[] $-\ E_{k,s}$: SoC at step $s$.
%     \item[] $-\ \alpha_{k,s}$: normalized FR signal $\in [-1,1]$.
%     \item[] $-\ E_{max}$: nominal battery capacity, reduced by $(1 - c_f)$ to reflect the remaining usable energy.
%     \item[] $-\ \lambda_l, \lambda_u, \epsilon_l, \epsilon_u \in (0,1)$: margins to ensure energy reserves.
%     \item[] $-\ c_f$: Cumulative capacity fade.
% \end{itemize}

Constraints (1.1) to (1.6) enforce the battery dynamics and operating limits. Equation (1.1) updates the SoC. Equation (1.2) limits the charging and discharging power to be within the rated bounds. Equation (1.3) restricts the SoC within allowable margins, accounting for usable capacity reduction due to degradation. Equation (1.4) enforces the power balance. Equation (1.5) enforces non-negativity on the decision variables. Finally, equation (1.6) imposes terminal constraints on the SoC to ensure feasibility across horizons.

The objective is to maximize short-term profit by valuing FR capacity revenues against the cost of grid purchases. Battery degradation is not explicitly optimized here, since the simplified model cannot capture it. Instead, MPC actions are later simulated in the high-fidelity DAE model \citep{santhanagopalan2006review} to evaluate degradation metrics for benchmarking. %Following prior work \citep{li2022deep}, we assume that the battery does not purchase power from the grid and discharges to the load at the same time by enforcing $O_t\cdot L_t=0$. 
We generate the optimal state–action pairs dataset by solving optimization Problem (1) in a low-fidelity environment once per hour until the battery reaches its EOL threshold to produce a batch of state–action pairs for pretraining.
%\vspace{-2pt}
\subsection{Behavior Cloning}

With the state-action pairs generated from the previous stage, we can approximate the MPC behavior by training a policy via BC. The aim is to imitate the behavior of a simple controller and provide a stable initialization for subsequent RL training.
%Accordingly, a deep neural network (DNN) is trained to map states to MPC actions, which will later be used for the initialization of RL in Section \ref{sec: degradation-aware-RL-training}. The policy approximation is expressed as follows:
The actor network approximates the control policy $\mu_{\boldsymbol{\theta}}(\mathbf{x_t}) = f(\mathbf{x_t}; \boldsymbol{\theta})$ where $\mathbf{x_t} = [\mathrm{mean}(\alpha_{k,s}),\ \mathrm{var}(\alpha_{k,s}),\ \pi_k^f,\ \pi_k^e,\ E_{k,t},\ C_t^f]^T \in \mathbb{R}^{n_x}$ 
is the state vector, and 
$\boldsymbol{\theta}\in \mathbb{R}^{n_\theta}$ 
are parameters of the Deep Neural Network (DNN). 
%  = [W^{(1)},\ b^{(1)},\ \ldots,\ W^{(L)},\ b^{(L)}]^T
%, consisting of weights $W^{(l)}$ and biases $b^{(l)}$ for each of the $L$ layers. 
The network is trained to reproduce the MPC actions $\mathbf{u^{MPC}_t} = [F_t,\; O_t - L_t]^T$ by minimizing a mean-squared error objective:
%\vspace{-5pt}
%The DNN outputs, $\mathbf{y_i}$, are the committed FR band, $F_t$, and the net grid decision, $O_t - L_t$. 
%Therefore, the SL problem can be written as follows:

\begin{equation}
\min_{\boldsymbol{\theta}} \frac{1}{M} \sum_{i=1}^M \left\| f(\mathbf{x}_i, \boldsymbol{\theta}) - {\mathbf{u}_i}^\mathbf{{MPC}} \right\|,
\end{equation}

where $M$ is the number of state-action samples in the batch.

In addition, we trained a critic network $Q_{\boldsymbol{\phi}}(\mathbf{x}_t, \mathbf{u}_t)$ to estimate the long-term value of each action. Its parameters are initialized using temporal-difference (TD) targets based on the MPC rollouts:
%We model the critic as:

%\begin{equation}
%Q_{\boldsymbol{\phi}}(\mathbf{x}_t, \mathbf{u}_t) = g(\mathbf{x}_t, \mathbf{u}_t; \boldsymbol{\phi})
%\end{equation}

%Here $\boldsymbol{\phi} = [W^{(1)},\ b^{(1)},\ \ldots,\ W^{(L)},\ b^{(L)}]^T$ are the trainable weights and biases. 
%Using the MPC dataset, we train $Q_{\boldsymbol{\phi}}$ with a temporal-difference (TD) target, 
%which combines the immediate reward from the current step with the estimated value of the next state–action pair:

\begin{equation}
\mathbf{y}_i = r_i + \gamma Q_{\boldsymbol{\phi}^-}(\mathbf{x}_i', \mathbf{u}_i'),
\end{equation}

where $r_i$ is the reward of the current decision, $\gamma \in (0,1)$ is the discount factor, and 
$Q_{\boldsymbol{\phi}^-}(\mathbf{x}_t', \mathbf{u}_t')$ is the critic’s estimate of the value of the next state–action pair 
under the target network $\boldsymbol{\phi}^-$. 

%For stable learning, the target network parameters $\boldsymbol{\phi}^-$ are maintained as a slower-moving copy of the critic’s parameters $\boldsymbol{\phi}$. After each update, $\boldsymbol{\phi}^-$ is moved slowly toward $\boldsymbol{\phi}$ based on a soft update rule.

The critic parameters are then optimized to minimize the differences between its predicted value and the TD target:

\begin{equation}
\min_{\boldsymbol{\phi}} \frac{1}{M} \sum_{i=1}^M \left\| Q_{\boldsymbol{\phi}}(\mathbf{x}_i, \mathbf{u}_i) - \mathbf{y}_i \right\|.
\end{equation}

\subsection{Degradation-Aware RL Training}
\label{sec: degradation-aware-RL-training}
%RL enables sequential decision-making in uncertain and fast dynamic systems. 
At each time step $t$, the agent observes the current state 
$\mathbf{x}_t$ and selects an action $\mathbf{u}_t = [F_t,\ O_t - L_t]^T \in \mathbb{R}^m$ according to its policy $\mu_{\boldsymbol{\theta}}(\mathbf{x}_t)$, 
and interacts with the environment and receives a reward $r_t(\mathbf{x}_t, \mathbf{u}_t, \mathbf{x}_{t+1}) \in \mathbb{R}$ and 
the next state $\mathbf{x}_{t+1}$ from the environment. 

The actor represents the policy $\boldsymbol{\mu}_{\boldsymbol{\theta}}(\mathbf{x}_t)$, which provides the deterministic action for state $\mathbf{x}_t$ (with exploration noise added during training), while the critic estimates 
the value of state–action pairs, $Q_{\boldsymbol{\phi}}(\mathbf{x}_t, \mathbf{u}_t)$, guiding the actor toward better decisions. 
The RL agent is initialized using actor and critic networks pretrained with SL to reduce sample complexity and avoid poor early-stage behavior.

%The Deep Deterministic Policy Gradient (DDPG) algorithm has been used for RL training, which extends deterministic policy gradient methods with deep neural networks. 

The Deep Deterministic Policy Gradient (DDPG) algorithm \citep{lillicrap2019} has been used for RL training, which is applicable here as the action space is continuous and bounded, satisfying standard conditions for deterministic policy gradient methods. The agent’s objective is to maximize the expected cumulative reward over an infinite horizon. When the action is generated by the policy, $\mathbf{u}_t = \mu_{\boldsymbol{\theta}}(\mathbf{x}_t)$
the reward expression simplifies to $r(\mathbf{x}_t, \mu_{\boldsymbol{\theta}}(\mathbf{x}_t))$ and the objective can be written as: 

\begin{equation}
J(\mu_{\boldsymbol{\theta}}) = \mathbb{E} \left[ \sum_{t=0}^{\infty} \gamma^t r(\mathbf{x}_t, \mu_{\boldsymbol{\theta}}(\mathbf{x}_t)) \right],
\end{equation}

where $\gamma$ is the discount factor that balances immediate and future rewards and  $\mathbb{E}[.]$ is the expected value defined over both the transition dynamics and the stage rewards. The reward function is expressed as:

\begin{equation}
r_t = 
\underbrace{\pi^{f_t} F_t - \pi^{e_t} O_t}_{\text{operational profit}} 
- 
\pi^{c_f} \underbrace{\left(C_{t+1}^f - C_t^f\right)}_{\text{fade penalty}},
\end{equation}

where the first term is operational profit, and the second term penalizes the battery capacity fade. The parameter $\pi^{C_f}$ reflects the long-term economic value of battery capacity and is set to \$ 12,000 \citep{cao2020multiscale}. This formulation includes a degradation penalty that makes the reward function degradation-aware. Using a high-fidelity environment enables this, as it explicitly captures battery degradation dynamics. To emulate real measurement conditions, a small random noise was added to the SoC values during training. %The degradation-aware agent is only used during training. The best policy actions serve as the reference setpoints ($F^{ref}_t$ and $O^{ref}_t$) used in Section \ref{sec:degradation-unaware-RL} and it is not deployed at runtime.

\subsection{Degradation-Unaware RL training}
\label{sec:degradation-unaware-RL}

The low-level controller must act at the 2-second FR timescale and track the MPC-generated setpoints in realtime. The degradation-aware agent from Section \ref{sec: degradation-aware-RL-training} is not suited for this role as its state includes DAE-derived variables (SEI thickness, capacity fade) that are too expensive to compute online, and it was trained for setpoint tracking. The degradation-unaware agent is therefore fine tuned from the degradation-aware weights via anchor-based refinement. The pretraining of the RL with a degradation-unaware setting uses the same DDPG framework and actor–critic setup described in Section \ref{sec: degradation-aware-RL-training}. However, the reward function does not include the degradation penalty term and is expressed as:
%\vspace{-2pt}
% \begin{equation}
% r_t = \pi_t^f F_t - \pi_t^e O_t - \lambda_{sp} \left[ (F_t - F_t^{ref})^2 + (O_t - O_t^{ref})^2 \right],
% \label{eq:8}
% \end{equation}

\begin{equation}
r_t = \underbrace{\pi^f_t F_t - \pi^e_t O_t}_{\text{operational profit}} 
      - \underbrace{\lambda_{\text{sp}}\Big[(F_t - F_t^{\text{ref}})^2 + (O_t - O_t^{\text{ref}})^2\Big]}_{\text{setpoint-tracking penalty}}
\label{eq:reward_du}
\end{equation}

where $F_t^{ref}$ and $O_t^{ref}$ are the reference setpoints for FR band and grid purchase, respectively. These references are obtained from the actions of the best policy trained in Section \ref{sec: degradation-aware-RL-training}. This reward function corresponds to the operational profit from FR revenues minus grid purchase costs, along with a penalty for deviation from the provided setpoints weighted by $\lambda_{sp}$. 

The degradation-unaware agent serves as the low-level controller that receives setpoints, $F_t^{ref}$  and $O_t^{ref}$, from the upper-level MPC and tracks them while maximizing operational profit. To explicitly improve setpoint tracking, the state vector was extended to include setpoints and their errors. The new state representation is: 
%\vspace{-2pt}
\begin{equation}
\begin{split}
\mathbf{x}_t = [F_t^{ref}, O_t^{ref}, e_t^{F}, e_t^{L}, \text{mean}(a_{k,s}),\\
\text{var}(a_{k,s}), \pi_k^{f}, \pi_k^{e}, E_{k,t}, C_t^{f}]^{T}.
\end{split}
\label{eq:9}
\end{equation}

Since the pretrained actor and critic from Section \ref{sec: degradation-aware-RL-training} were designed with only six state inputs, we reused their learned weights and biases as initialization and expanded the input layers to be able to accept the additional setpoints. The networks were then retrained under the new reward function. SoC safety is enforced hierarchically. First, the MPC solves constraints (1.2)-(1.3) at the slow timescale, producing feasible setpoints $F_t^{\text{ref}}, O_t^{\text{ref}}$ that respect $(1-c_f)\lambda_l E_{\max} \leq E \leq (1-c_f)\lambda_u E_{\max}$. Second, the degradation unaware-RL agent is trained with a setpoint-tracking penalty $\lambda_{\text{sp}}[(F_t-F_t^{\text{ref}})^2 + (O_t-O_t^{\text{ref}})^2]$ in Equation \ref{eq:8}, which implicitly keeps actions within the MPC-safe envelope. Third, at runtime, RL outputs are projected onto $[P_{\min}, P_{\max}]$ and clipped to the MPC SoC bounds. This three-layer guarantee ensures constraint satisfaction even under model mismatch or exploration noise.%SoC safety is guaranteed at the MPC level through constraints (1.2)-(1.3), scaled by the degradation factor (1-$c_f$). Since the RL agent is trained to track these MPC-generated setpoints, which already respect the safe operating bounds, constraint satisfaction is preserved at the lower level as a consequence of the hierarchical design.

We used anchor-based refinement for the critic to avoid aggressive shift from the pretrained model while still allowing adaptation to the new objective. This mechanism anchors the new critic $Q_{\boldsymbol{\phi}}'$ to the old one $Q_{\boldsymbol{\phi}}$ by penalizing large deviations:
%\vspace{-2pt}
\begin{equation}
\begin{split}
L(\boldsymbol{\phi}) = &\, \frac{1}{M} \sum_{i=1}^M \left( Q_{\phi}'(\mathbf{x_i}, \mathbf{u_i}) - \mathbf{y_i} \right)^2 \\& + \lambda_a \left\| Q_{\phi}'(\mathbf{x_i}, \mathbf{a_i}) - Q_{\phi}(\mathbf{x_i}, \mathbf{u_i}) \right\|^2,
\end{split}
\end{equation}

where $\mathbf{y_i}$ is the TD target and the second term enforces anchoring to the previously trained critic with anchoring weight $\lambda_a>0$.

\subsection{Hierarchical Control Strategy}
\label{sec:hierarchical-control-strategy}
In hierarchical control strategy, we break down a complex decision problem into layers that operate on different timescales and levels of detail. The higher layer focuses on strategic goals, while the lower layer executes more detailed and faster actions \citep{olanrewaju2016economic}. Furthermore, the hierarchical framework adds modularity and improves learning efficiency even under uncertain and nonlinear battery dynamics such as high-fidelity environment. %Building on this idea, we developed a hierarchical control strategy with two layers.
At the high level, an MPC operates on a slower time scale and optimizes degradation-aware setpoints for the FR band and grid power over a prediction horizon. 
%At the low-level layer, a pretrained RL agent runs at a faster resolution, executing actions that maximize operational profit while tracking the setpoints generated by the high-level controller. Fig. \ref{Fig. 2} shows the schematic of this proposed strategy.
% \begin{figure}[t!]
%     \centering
%     \includegraphics[width=0.7\linewidth]{Figure2.png}
%     \vspace{-5pt}
%     \caption{MPC–RL hierarchical scheme. MPC serves as the higher-level controller and RL as the lower-level controller. RL received states [mean(signal), var(signal), FR price, grid price, SoC, capacity fade], whereas MPC received high-fidelity states [electrode concentrations, electrode potentials, currents, SEI variables, capacity fade].}
%     \vspace{-15pt}
%     \label{Fig. 2}
% \end{figure}

We used a high-fidelity battery environment that better reflects real-world degradation and improves the prediction of long-term degradation trends. However, solving MPC in the high-fidelity environment is computationally expensive. To address this, the MPC uses a 20-minute time step and a prediction horizon for 4 hours, which limits long-term anticipation of degradation but maintains the optimization tractable. %we made two simplifications: (1) increased the simulation time step to 20 minutes to reduce the number of optimization steps; (2) limited the prediction horizon to 4 hours to keep the number of variables in each optimization manageable.

%However, these simplifications can limit the controller’s long-term foresight, but make optimization tractable and feasible.

The MPC objective function is defined as follows:
 %\vspace{-10pt}
 
\begin{equation}
\max_{\{F_k, O_k, L_k, P_{k,s}, E_{k,s}\}} \sum_{k=t}^{t+N-1} \epsilon  \left( \pi_k^f F_k - \pi_k^e O_k \right) - \pi^{c_f} \Delta C_k^f,
\end{equation}

where $\epsilon\ll1$ is a small revenue weight, $\Delta C_k^f$ is the capacity loss caused by battery degradation. This objective function ensures that degradation minimization is the primary objective of the high-level controller, while still allowing limited participation in the market to avoid trivial zero setpoints. All terms are in their physical units without normalization.In the low-level controller, we implemented the pretrained RL actor from Section \ref{sec:degradation-unaware-RL}. 

\subsection{Problem set-up}

The simulation environment was built on the work of  \citep{cao2020multiscale}, which we adapted and extended for our framework. All coding was carried out in Julia.

For the MPC, the optimization problem is modeled using the JuMP package and solved with the GLPK solver. For the SL stage, the policy and value networks were implemented in Flux and trained with the ADAM optimizer over 5000 epochs. The policy network had four layers with 10, 30, 15, and 2 neurons, using ReLU activation in the hidden layers and tanh in the output layer. The critic network also had four layers with 12, 30, 15, and 1 neurons, with ReLU activation throughout. For the RL training setup, one week was taken as an episode because electricity prices follow a weekly cycle. To encourage exploration, Gaussian noise was added to the actor outputs for better training. In degradation-unaware RL, the setpoint tracking weight was set to $\lambda_{sp}=0.2$. The DDPG training used $\gamma = 0.9$, a batch size of 160, an exploration noise scale of 0.05, and an anchor weight of 0.5. Finally, in the hierarchical MPC–RL framework, the upper-level MPC was solved using Ipopt.

We used three different battery setup, LFP, NMC and NCA  to test generalizability of the framework and RL agent. The maximum charge and discharge were set to 10 MW, and the number of cells was scaled to 1 MWh capacity. We compare the performance of developed FR strategies (Section \ref{sec:mpc} and Section \ref{sec:hierarchical-control-strategy}). %the MPC derived from Section 2.3, and the MPC-RL: the hierarchical control framework developed in Section 2.7 
Each FR policy was simulated until the battery reached its EOL, which was defined as a 20\% capacity fade. 

\section{Results and Discussions}
%\vspace{-5pt}
Fig. \ref{Fig. 3} shows the operational profit of battery overtime for different FR policies under different chemistries. The trajectories differ in length because each simulation was run until the battery reached its EOL. The performance of the implemented FR policies has a significant effect on the battery longevity and profit. MPC alone has poor performance, which indicates that its actions quickly exhaust the battery, resulting in lower cumulative operational profit and faster degradation compared to MPC-RL. The extended battery lifetime under the MPC-RL policy enables longer market participation and results in higher cumulative operational profit. While MPC–RL performed better, it required higher computational effort, averaging 15 s per action compared to 6 s for MPC. This overhead stems from the additional coordination between MPC and RL layers and high-fidelity environment.

Moreover, the consistent performance of the MPC-RL policy across different battery types demonstrates the generalizability of the pretrained RL agent. Despite the varying electrochemical characteristics of each chemistry, the MPC-RL controller successfully adapts its actions to maintain stable profitability trends and slower degradation rates.

\begin{figure}[ht!]
    \centering
    \includegraphics[width=0.95\linewidth]{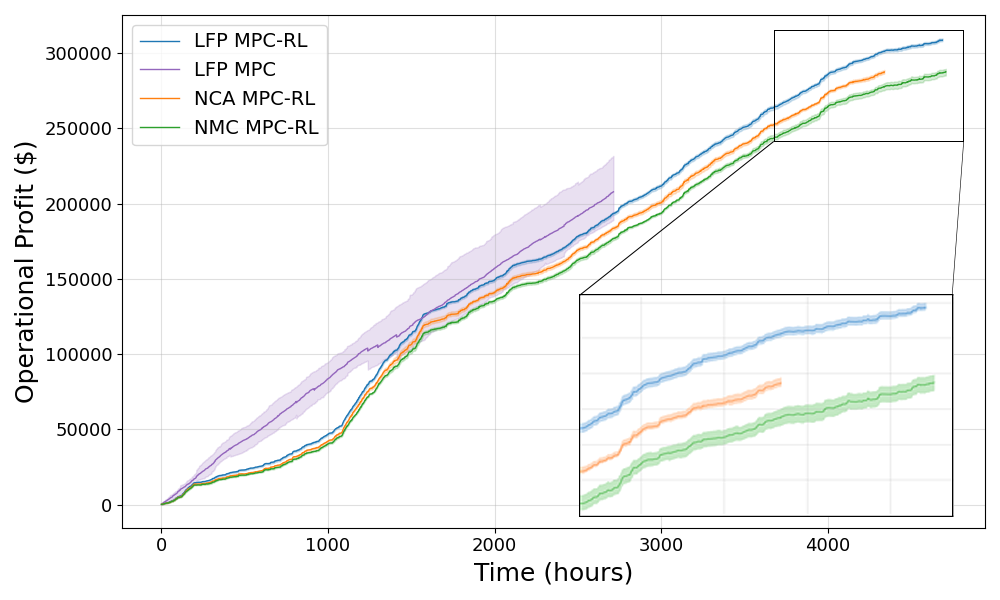}
    \vspace{-5pt}
    \caption{Operational Profit overtime for different FR policies under different chemistries.}
    
    \label{Fig. 3}
\end{figure}

\begin{figure}[ht!]
    \centering
    \includegraphics[width=1.0\linewidth]{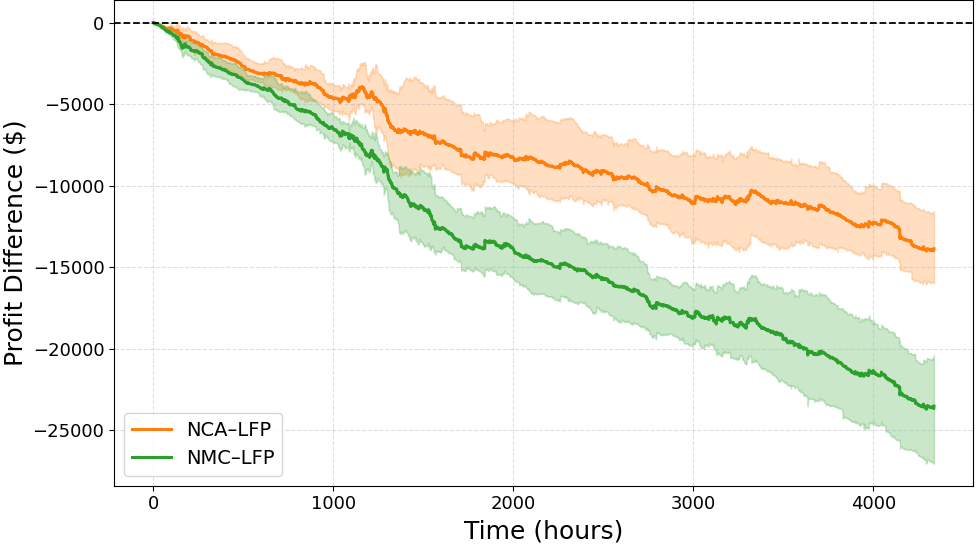}
    \vspace{-10pt}
    \caption{Profit difference of NCA and NMC batteries relative to the LFP MPC–RL baseline over the battery lifetime.}
    \label{Fig. 4}
\end{figure}

 Fig. \ref{Fig. 4} shows the difference in operational profit of NCA and NMC batteries compared to LFP over time. Both NCA and NMC perform worse than LFP, and the gap continues to increase as the batteries age. NCA performs slightly better, but still trails behind LFP, while NMC falls much further behind.

In addition, maintaining the SoC within the range of 0.1 – 0.9 is essential for efficient operation, and this constraint was enforced across all FR policies. Table \ref{Table 1} shows how each policy managed SoC within this range. In both cases, the SoC remained almost entirely within the safe window of 0.2 to 0.8, with very few excursions. These results suggest that MPC-RL can effectively maintain constraints.

\begin{table}[h!]
\centering

\caption{Statistical summary of SoC under different FR control policies.}
\label{Table 1}
\resizebox{0.95\columnwidth}{!}{
\begin{tabularx}{\columnwidth}{
    l
    *{4}{>{\centering\arraybackslash}X}
    >{\centering\arraybackslash}X
    @{\hspace{2em}}
    >{\centering\arraybackslash}X
}
\hline
\makecell{FR \\Policy} & 
\makecell{Mean \\SoC} & 
\makecell{Std. \\Dev \\SoC} & 
\makecell{Min \\SoC} & 
\makecell{Max \\SoC} & 
\makecell{\%Time\\SoC\\[0.2ex]$<0.2$} &
\makecell{\%Time\\SoC\\[0.2ex]$>0.8$} \\
\hline
LFP MPC-RL & 0.42 & 0.06 & 0.13 & 0.87 & 0.21 & 0.13\\
LFP MPC    & 0.46 & 0.12 & 0.15 & 0.90 & 0.30 & 2.25\\
NCA MPC-RL & 0.39 & 0.08 & 0.11 & 0.74 & 0.57 & 0.00\\
NMC MPC-RL & 0.39 & 0.08 & 0.11 & 0.70 & 0.99 & 0.00\\
\hline
\end{tabularx}}
\end{table}

\begin{figure*}[h!]
    \centering
    \includegraphics[width=0.95\textwidth]{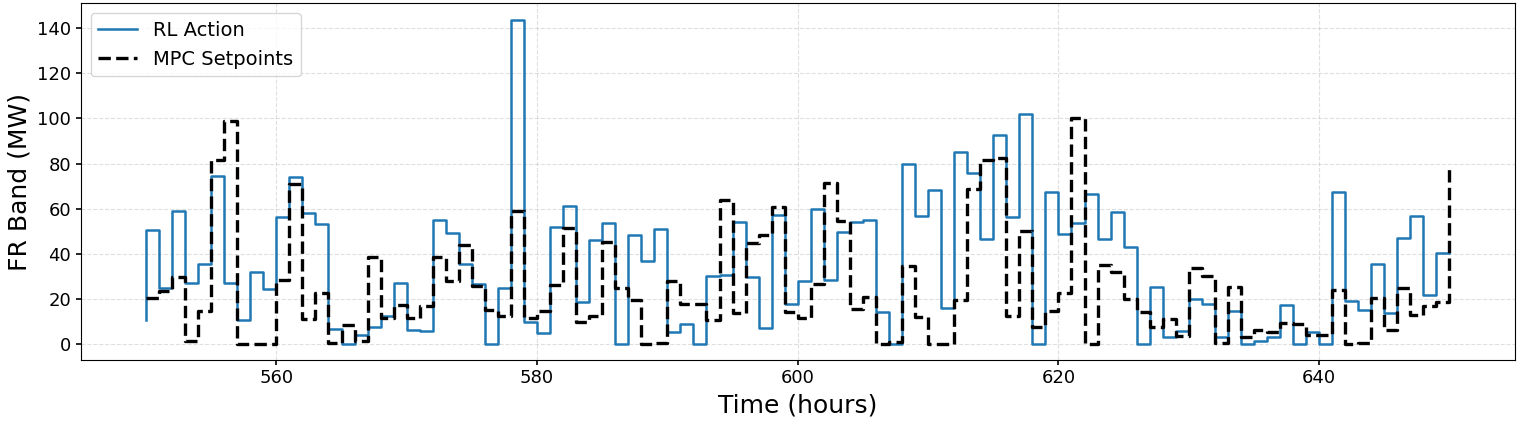}
    \caption{MPC–RL setpoint tracking over a sample period. The RL controller occasionally pushes the FR band above the MPC target to capture short-term profit.}
    \label{Fig. 5}
\end{figure*}

Fig. \ref{Fig. 5} shows how closely RL actions follow the MPC-generated setpoints over a sample period. In the hierarchical control strategy, it is important to examine whether the RL actions follow the MPC-generated setpoints. Although some deviations are expected, especially during high-profit opportunities, the RL controller successfully captures the overall trends of the MPC setpoints, with only small deviations. The high correlation of 0.90 confirms the alignment between the RL actions and the MPC generated setpoints.

The results show that degradation-aware strategies bring significant improvements in both battery’s life and operational profit. The performance of the developed hierarchical strategy shows that the approach works effectively. By separating slow dynamic of battery degradation from fast market response decisions, the framework is able to handle multi-scale systems. This separation improves information use, with MPC focusing on battery health and setting safe operating boundaries, while RL focuses on market and short-term profit. Compared to MPC, MPC–RL extends battery life by roughly 84\% and increases operational profit by 34\% (about \$ 80,000). 

\vspace{-6pt}
\section{Conclusion}
\vspace{-5pt}
This study presented a novel hierarchical control strategy for the participation of BESS in FR markets to balance long-term degradation control with short-term revenue. The results show that degradation-aware methods clearly outperform MPC in both lifetime and revenue.%Quantitatively, MPC–RL extended battery life by 84\%, increased cumulative profit by approximately \$ 80,000, achieved 0.90 setpoint-tracking correlation, and generalized across three chemistries 
The MPC–RL framework was able to extend battery life by 84\% compared to MPC. It also increased operational profit by approximately \$ 80,000. %Although it is slightly more computationally expensive, it remains tolerable and feasible for practical use.
%The RL agent in the proposed framework was developed through a staged pretraining process. The setpoint-tracking capability was evaluated with 0.90 correlation between RL actions and MPC setpoints, confirming the reliability of the design. 
%Furthermore, the generalizability of the framework was tested in three different battery systems. Overall, the framework has the potential to be implemented in other multi-scale systems where objectives have different timescales. 

Despite its advantages, the framework has some limitations. Reliance on repeated nonlinear MPC optimization increases the computational cost, and simplifications in the high-fidelity model may limit generalizability.

For future work, the computational cost could be reduced by using surrogate models for the high-fidelity MPC or adaptive horizon strategies. The framework could also be extended to multi-service operation, where BESS participates in energy and ancillary service markets. At the algorithmic level, extensions such as multi-agent RL or multi-layer hierarchical structures could further improve coordination and scalability across other services.

\section*{Data and Code Availability}
\vspace{-5pt}
All data and code used in this study are available at the following GitHub repository:

\url{https://github.com/rasaprj/FR_market/tree/Hierarchical_MPC-RL}.
\vspace{-5pt}
\begin{ack}
%\vspace{-10pt}
The authors acknowledge the use of the High-Performance Computing facilities provided by Imperial College London to support this work.
%\vspace{-5pt}
\end{ack}

\section*{DECLARATION OF GENERATIVE AI AND AI-ASSISTED TECHNOLOGIES IN THE WRITING PROCESS}
%\vspace{-10pt}
During the preparation of this work, the authors used ChatGPT (OpenAI, San Francisco, CA) in order to improve language clarity and grammar. After using this tool/service, the authors reviewed and edited the content as needed and take full responsibility for the content of the publication.

%\vspace{-10pt}

\bibliography{ifacconf}             % bib file to produce the bibliography

\end{document}